\documentclass[journal]{IEEEtran}

\usepackage{myStyleIEEE}
\usepackage{nomencl}
\usepackage{ragged2e}
\IEEEoverridecommandlockouts

\newlength{\bracewidth}
\newcommand{\myunderbrace}[2]{\settowidth{\bracewidth}{$#1$}#1\hspace*{-1\bracewidth}\smash{\underbrace{\makebox{\phantom{$#1$}}}_{#2}}}

\usepackage{etoolbox}
\renewcommand\nomgroup[1]{%
  \item[\bfseries
  \ifstrequal{#1}{P}{A. Parameters}{%
  \ifstrequal{#1}{V}{C. Variables}{%
  \ifstrequal{#1}{S}{B. Sets and Indices}{}}}%
]}

\begin{document}

\title{Mitigating Load-Altering Attacks Against Power Grids Using Cyber-Resilient Economic Dispatch}

\newtheorem{proposition}{Proposition}
\renewcommand{\theenumi}{\alph{enumi}}

\author{Zhongda~Chu,~\IEEEmembership{Student~Member,~IEEE,} 
        Subhash Lakshminarayana,~\IEEEmembership{Senior Member,~IEEE}\\
        Balarko Chaudhuri,~\IEEEmembership{Senior Member,~IEEE} and
        Fei~Teng,~\IEEEmembership{Senior Member,~IEEE} 
        \thanks{

        Zhongda Chu, Balarko Chaudhuri and Fei Teng are with Department of Electrical and Electronic Engineering, Imperial College London, London SW7 2AZ, U.K. Corresponding author:  Dr  Fei  Teng (f.teng@imperial.ac.uk).
        
        Subhash Lakshminarayana  is with the School of Engineering, University of Warwick, Coventry CV4 7AL, U.K.
        } 
        
\vspace{-0.5cm}}
\maketitle
\IEEEpeerreviewmaketitle

\begin{abstract}
Large-scale Load-Altering Attacks (LAAs) against Internet-of-Things (IoT) enabled high-wattage electrical appliances (e.g., wifi-enabled air-conditioners, electric vehicles, etc.) pose a serious threat to power systems' security and stability. In this work, a Cyber-Resilient Economic Dispatch (CRED) framework is presented to mitigate the destabilizing effect of LAAs while minimizing the overall operational cost by dynamically optimizing the frequency droop control gains of Inverter-Based Resources (IBRs). The system frequency dynamics incorporating both LAAs and the IBR droop control are modeled. The system stability constraints are explicitly derived based on parametric sensitivities. To incorporate them into the CRED model and minimize the error of the sensitivity analysis, a recursive linearization method is further proposed. A distributionally robust approach is applied to account for the uncertainty associated with the LAA detection/parameter estimation. The overall performance of the proposed CRED model is demonstrated through simulations in a modified IEEE reliability test system.
\end{abstract}


\makenomenclature
\renewcommand{\nomname}{List of Symbols}
\mbox{}
\nomenclature[P]{$M$}{system inertia$\,[\mathrm{MWs/Hz}]$}

\section{Introduction} \label{sec:1}
Internet-of-things (IoT)-enabled electrical appliances, such as WiFi-enabled air conditioners, electric vehicles, etc. are being increasingly installed, as they provide convenience to consumers. However, these devices typically do not have sophisticated security measures (such as encryption-enabled communication) incorporated in their design. Thus, they may become convenient entry points for malicious actors to gain access to the power grid. Botnet-type attacks that target a large number of these devices can severely affect power grid operations.

The topic of power grid security has gained
widespread prominence in the last decade. The majority of the works focus on the security of bulk power and the associated supervisory control and data acquisition (SCADA) system \cite{Liu2009, 9756414, LakshDataDrive2021}. In contrast, cyber attacks targeting end-user IoT-electrical appliances have only gained attention recently. Existing work on this topic can be broadly divided into two categories -- (i) attack impact analysis, and (ii) detecting/localizing cyber attacks.

An unplanned and large-scale fluctuation in the system load can potentially disrupt the balance between supply and demand, subsequently leading to unsafe operating conditions. LAAs can be divided into categories depending on the type of load altered -- \emph{static} LAAs and \emph{dynamic} LAAs. Static LAAs refer to a sudden one-time manipulation of the demand. It has been shown that such attacks lead to economic cost increment, load shedding, and/or unsafe frequency excursions \cite{Dabrowski2017, soltan2018blackiot, HuangUSENIX2019}. Temporal fluctuations of system load and renewable energy can be exploited by the attacker to execute LAAs \cite{Ospina2021, LakshCOVID2022}. Furthermore, low-inertia conditions caused by the growing penetration of renewable energy resources can further exacerbate the vulnerability of power grids to LAAs \cite{LakshCOVID2022}. On the other hand, Dynamic LAA (DLAA) involves the manipulation of the demand over several time slots. In particular, it was shown that if attackers can manipulate the load in accordance with the fluctuations of system frequency, they can potentially destabilize the frequency control loop  \cite{AminiLAA2018}. Moreover, the knowledge required to execute such attacks can be gathered from publicly-available information \cite{AcharyaPHEV2020}.
An analytical framework to identify buses corresponding to the least-effort destabilizing attacks was characterized in \cite{LakshIoT2021} using the theory of second-order dynamical systems. 


Detecting and localizing LAAs is an important first step towards mitigating their destabilizing effects.
Initial work in this area aimed at detecting LAAs by analyzing the power grid signals (phase angle/frequency data) in the frequency domain \cite{AminiDetect2015}, and this method was subsequently extended to localizing the nodes that are under attack \cite{AminiIdentification2019}. However, the framework was restricted to a linear model of power system dynamics, and cannot be extended trivially to non-linear models. An unscented Kalman filter-based approach to detect and localize LAAs was proposed in \cite{IzbickiACC2017}, which, however, does not scale well for large power grids. To overcome these limitations, reference \cite{lakshminarayana2021datadriven} proposed the application of physics-informed machine learning techniques to detect and localize LAAs by monitoring data from phasor measurement units following the attack. The proposed framework can also estimate the attack parameters (static attack magnitude and attack controller gain for dynamic attack).

In contrast to the two streams of work, the problem of mitigating the destabilizing effects of LAAs and the corresponding impact on system operation has received limited attention, which is the main focus of this paper. Existing works on this topic mostly focus on offline security enhancements. For instance, the problem of finding the optimal locations for deploying load protection features (e.g., encryption-enabled smart devices) has been addressed in  \cite{AminiLAA2018, LakshIoT2021}. Reference \cite{SoltanTNSE2020} formulates an optimization framework to obtain generator operating points to ensure that none of the transmission lines in the grid become overloaded in the event of an LAA. A data-centric edge-computing infrastructure for IoT devices in power grids is designed in \cite{9089080}. Based on power grid knowledge, edge servers are equipped to enhance the security policies against IoT-based attacks. A model-free defense framework against LAAs, using machine-learning based algorithms is proposed in \cite{CHEN2020116015} where only static LAAs are considered. Reference \cite{GUO2021107113} presents a defense policy against LAAs, which models the problem through a two-player zero-sum multistage game solved by minimax-q learning.

However, such offline mitigation measures incur significant economic costs due to two reasons. (i) The system must be operated at an uneconomic point over the entire operating period, irrespective of whether the attack occurs or not (cyber attacks are very rare events). (ii) Since the operator has no way of having prior knowledge of the actual parameters the attackers may use, they require making \emph{worst-case} assumptions about the attacker's capabilities (e.g., secure the power grid against all possible attack parameters).

To address these shortcomings, in this work we propose an online DLAA mitigation technique to enhance the power system's resilience to potential LAAs, named, cyber-resilient economic dispatch (CRED). The proposed method utilizes the fast and flexible control of IBR units, whose capability to enhance the system security and stability has been extensively studied and demonstrated such as in \cite{9329077,8743441,9475967,8506376}. Specifically, since the DLAAs destabilize the system by decreasing the system damping, additional damping achieved through IBR droop control is supplied to the grid when an attack event is detected. \textcolor{black}{It is to be noted that the scope of this study is limited to traditional electromechanical stability. The assumption is that IBRs are controlled to provide synthetic inertia and mimic the operation of SGs. Hence, the basic nature of the stability problem remains electromechanical even with large penetration of IBRs. Stability problems caused by IBRs operating in grid-following mode in conjunction with a weak network \cite{9286772} in particular, are outside the scope of this study. The droop gains of the IBRs are known to influence such stability problems which would be accounted for in a future study. Additionally, the control and response of IBRs are much faster and more accurate than SGs, hence enabling the IBR units to provide the desired damping to the grid \cite{6683080,8579100}. Although it is possible to design or utilize more sophisticated IBR control schemes, it would significantly complicate the system dynamics and prevent it from being incorporated into optimization model, thus not being considered in this work.}
Moreover, the proposed approach explicitly takes the attack detection/localization results into account, including the uncertainty associated with them. It can also be easily incorporated into the conventional UC problem. Our main contributions include the following:
\begin{itemize}
    \item We propose a CRED model to ensure the power system stability under potential LAAs. The proposed scheduling framework is implemented dynamically following LAA detection/identification to achieve stable system operation until the attack is eventually isolated in the most cost-effective manner.
    \item We derive the system stability constraints analytically based on parametric sensitivities. To improve the accuracy of eigenvalue sensitivities, a recursive linearization approach is proposed. The resulting constraints are further reformulated while considering the uncertainty associated with the attack parameter estimation, to fit the overall optimization model.
    \item The performance and effectiveness of the proposed model are demonstrated through case studies based on an IEEE Reliability Test System. The cyber-resilience improvement under different operating conditions and the corresponding cost increment as well as the impact of various factors are assessed.
\end{itemize}
The rest of the paper is organized as follows. Section \ref{sec:2} introduces the power system model considering the LAAs and the mitigation strategies. Section \ref{sec:3} derives system stability constraints, which are reformed to decrease the sensitivity error and fit the optimization model. The uncertainty of attack detection and the overall operation framework are discussed in Section \ref{sec:4}, followed by case studies in Section \ref{sec:5}. Section \ref{sec:6} concludes the paper.

\section{System Model} \label{sec:2}
In this section, the frequency dynamic model in a multi-area power system, the LAA model and the proposed mitigating strategies against the attacks are discussed.

\subsection{Power System Modeling}
We consider a power system containing a set of $\mathcal{N}=\{1,...,N\}$ areas with the frequency and phase angle in each area assumed to be the same. Each area may include Synchronous Generators (SGs), IBRs, storage units and loads. The power system's phase angle and frequency dynamics can be described by the following set of differential equations:
\begin{align}
\label{sys_dyn0}
    \begin{bmatrix}
    I&O\\
    O & -\myunderbrace{(M_g+M_s)}{M}
    \end{bmatrix}  \begin{bmatrix}
    \dot \delta\\ \dot \omega 
    \end{bmatrix} &=\begin{bmatrix}
    0\\ {P^L-P^C} 
    \end{bmatrix} \nonumber \\
    &+
     \begin{bmatrix}
    O & I\\
    K^I+B & K^P + D
    \end{bmatrix} \begin{bmatrix}
     \delta\\  \omega 
    \end{bmatrix} ,
\end{align}
where $\delta,\,\omega \in \mathbb{R}^{N}$ are vectors of phase angles and frequency deviations in each area respectively; $M_g,\,M_s,\,D \in \mathbb{R}^{{N}\times{N}}$ are diagonal matrices with their diagonal entries being the aggregated inertia of SGs, aggregated synthetic inertia from IBRs and the aggregated damping from SGs and loads; $K^I,\,K^P \in \mathbb{R}^{{N}\times{N}}$ are diagonal matrices with their diagonal entries being the aggregated integral and proportional gains of the governor control loop of the SGs in each area respectively; $B\in \mathbb{R}^{{N}\times{N}}$ denotes the susceptance matrix connecting different areas, and $P^L \in \mathbb{R}^{N}$ and $P^C \in \mathbb{R}^{N}$ represent the vector of aggregated demands and IBR outputs in each area.

\subsection{Load Altering Attacks}
Under LAAs, the attacker manipulates the system load by synchronously turning on or off a large amount of electrical IoT-enabled appliances. It is assumed that the demand in each area consists of two components $P^L = P^{LS} + P^{LV}$, where $P^{LS}$ denotes the secure part of the load and $P^{LV}$ denotes the vulnerable part of the load, which the attacker can manipulate. The system vulnerable load under LAAs is given by:
\begin{equation}
\label{Pl}
    P^{LV} = -K^{L}\omega + \epsilon^L,
\end{equation}
where $K^{L} \omega$ is the DLAA component that represents a time-varying load manipulation that follows the local frequency fluctuations \cite{AminiLAA2018} and $\epsilon^L$ is the SLAA component (one-time step-change manipulation); $K^{L} \in \mathbb{R}^{{N}\times{N}}$ represents the matrix of attack controller gains. It is assumed that the attacker can only monitor the frequency within the attack area, i.e., $K^L$ being diagonal. Combining \eqref{sys_dyn0} and \eqref{Pl} gives the power system dynamics under load altering attack:
\begin{align}
\label{sys_dyn1}
    \begin{bmatrix}
    I&O\\
    O & -M
    \end{bmatrix}  \begin{bmatrix}
    \dot \delta\\ \dot \omega 
    \end{bmatrix} & =\begin{bmatrix}
    0\\ {P^{LS}+\epsilon^L-P^C} 
    \end{bmatrix} \nonumber \\
    & +
     \begin{bmatrix}
    O & I\\
    K^I+B & K^P + D-K^L
    \end{bmatrix} \begin{bmatrix}
     \delta\\  \omega 
    \end{bmatrix} .
\end{align}
Using the dynamic LAA component, the attacker can alter the eigenvalues of the system indirectly by changing the elements of the matrix $K^{L}$, which is equivalent to modify the overall damping in the system. Thus, they can potentially destabilize the power system control loop \cite{AminiLAA2018}. Although an attack with larger $K^L$ would lead to a less stable system, the choice of $K^L$ is limited by the following constraint:
\begin{equation}
\label{K^L_Limit}
    K^L_{n,n} \omega^{\mathrm{max}}\le (P^{LV}_n-\epsilon_n^L)/2, \,\,\forall n\in \mathcal{N},
\end{equation}
where $K^L_{n,n}$, the $n$-th diagonal element is the attack gain in the $n$-th area and $\omega^{\mathrm{max}}>0$ is the maximum permissible frequency deviation in the system. This constraint requires that the maximum value of the load to be altered by the attacker in each area should be less or equal to amount of vulnerable load after removing the static load altering attack ($P^{LV}_n-\epsilon_n^L$). The factor 2 on the RHS is due to the fact that to destabilize the system, the amount of load that needs to be compromised must allow for both over and under frequency fluctuations before the system frequency exceeds $\omega^{\mathrm{max}}$.

\subsection{Mitigating Actions Against LAAs} 
The attackers have the capability to destabilize the system if the amount of the vulnerable load is large enough. Therefore, it is necessary for the system operators to maintain the system stability with proper measures given the potential LAAs. The mitigation strategies considered here is the frequency droop control implemented in the IBRs, due to their fast and accurate active power control. Furthermore, the droop structure is relatively simple and can be incorporated into the system scheduling process, enabling dynamic optimization and implementation immediately after the LAA detection. Moreover, the droop control is equivalent to providing system damping which is exactly what the attacker manipulates through the DLAA, thus being effective in counteracting the attack and stabilizing the system. Note that the droop control has been proposed and studied mainly for the purpose of frequency support after system disturbances. However, in this work the frequency droop control from IBRs is controlled to supply additional system damping during the attack period such that the overall system stability can be ensured. 

The output power from IBRs, $P^C$ can then be expressed as
\begin{equation}
\label{Pc}
    P^C = P^{C*} -K^{C} \omega,
\end{equation}
where $P^{C*}$ is the total power reference of IBRs in each area and $K^{C} \omega$ represent the power deviation according to the local frequency. $K^C \in \mathbb{R}^{{N}\times{N}}$ is a diagonal matrix with the entries being the aggregated droop gains in each area. Combining \eqref{sys_dyn0}, \eqref{Pl} and \eqref{Pc} gives the power system dynamics with LAAs and the droop response from IBRs. 
\begin{align}
\label{sys_dyn2}
    \underbrace{\begin{bmatrix}
    I&O\\
    O & -M
    \end{bmatrix}}_\mathcal{-A}  \begin{bmatrix}
    \dot \delta\\ \dot \omega 
    \end{bmatrix} & =\begin{bmatrix}
    0\\ {P^{LS}+\epsilon^L-P^{C*}} 
    \end{bmatrix} \nonumber \\
    & +
    \underbrace{\begin{bmatrix}
    O & I\\
    K^I+B & K^P + D-K^L+K^C
    \end{bmatrix}}_\mathcal{B} \begin{bmatrix}
     \delta\\  \omega 
    \end{bmatrix} .
\end{align}
It is evident that with appropriate selection of the droop control gains from IBRs given the estimation or prediction of the potential attacks \cite{lakshminarayana2021datadriven}, the system stability can be maintained. Remarkably, the droop response of IBRS are considered as decision variables due to their fast control whereas the proportional control gains ($K^P$) of synchronous generators are considered to be fixed.

\section{System Stability Constraint Derivation and Reformulation} \label{sec:3}
In this section, the system stability constraints under LAAs are derived based on system eigenvalues and sensitivity analysis. To ensure an accurate approximation, a recursive linearization approach is further proposed. The resulting conditional constraints are effectively reformulated to fit the overall optimization structure while considering the uncertainty associated with the attack gain estimation.

\subsection{System Stability Constraints}
Denote the eigenvalues of system \eqref{sys_dyn2} by $\lambda_i\in \mathbb{C}$ and the associated right and left eigenvector by $\mathbf{z}_j,\,\mathbf{y}_j\in \mathbb{C}^{2N}$. The system becomes unstable if there exists at least one eigenvalue $\lambda_i$, such that $\Re{(\lambda_i)}>0$. \textcolor{black}{Here, only stability constraint is considered but if the requirement is for the oscillations to settle within a stipulated time, the constraint boundary can be shifted to the left of the imaginary axis to ensure a minimum settling time.} Therefore, to ensure the system stability, the following constraint must hold:
\begin{align}
\label{f_sta}
    \Re{(\hat \lambda_{i})} & = \Re{(\lambda_i^0)} + \sum_{n\in\mathcal{N}}  \Re{\left(\frac{\partial\lambda_i^0}{\partial K^L_{n,n}}\right)}K^L_{n,n} \nonumber \\ 
    &+ \sum_{n\in\mathcal{N}}  \Re{(\frac{\partial\lambda_i^0}{\partial K^C_{n,n}})}K^C_{n,n} <0,\, \forall i = 1,..., 2N,
\end{align}
where $\hat \lambda_{i}$ is the estimated $i$-th eigenvalue using sensitivities and $\lambda_i^0$ is the eigenvalue of the system without the LAAs and the additional droop control from IBRs. The relationship $\Re{(\lambda_i^0)}<0, \,\forall i = 1,..., 2N$ must hold since the system is stable without attacks. $\frac{\partial\lambda_i^0}{\partial K^L_{n,n}}/\frac{\partial\lambda_i^0}{\partial K^C_{n,n}}$ are the eigenvalue sensitivities with respect to the $n$-th diagonal element in $K^{L}/K^{C}$ matrix. The analytical expressions for the sensitivity factors can be derived using the theory of second-order dynamical system \cite{LakshIoT2021}:
\begin{equation}
\label{sensitivity}
    \frac{\partial\lambda_i^0}{\partial K^L_{n,n}} = - \frac{\partial\lambda_i^0}{\partial K^C_{n,n}} = - \mathbf{y}_i^\intercal 
    \begin{bmatrix}
        \lambda_i^0 \frac{\partial \mathcal{A}}{\partial K^L_{n,n}} + \frac{\partial \mathcal{B}}{\partial K^L_{n,n}}
    \end{bmatrix} \mathbf{z}_i,
\end{equation}
where,
\begin{subequations}
\begin{align}
    \frac{\partial \mathcal{A}}{\partial K^L_{n,n}} &= O \\
    \frac{\partial \mathcal{B}}{\partial K^L_{n,n}} &= \begin{bmatrix}
        O &O\\ 
        O &-I_{n,n}
    \end{bmatrix}.
\end{align}
\end{subequations}
We define $I_{n,n}$ as a matrix whose $(n,n)^{\mathrm{th}}$ element is one if there is a DLAA in the $n$-th area, and all the others are zero. 

Moreover, the limits of $K^L_{n,n}$ and $K^C_{n,n}$ in an actual system have to be considered. The former is discussed in \eqref{K^L_Limit}, whereas for $K^C_{n,n}$, the following constraint applies:
\begin{subequations}
\label{K^C_lim}
\begin{align}
    P^{C*}_n + K^{C}_{n,n} \omega^{\mathrm{max}} &\le P^{C,\mathrm{max}}_n \\
    P^{C*}_n - K^{C}_{n,n} \omega^{\mathrm{max}} &\ge 0, 
\end{align}
\end{subequations}
where $P^{C,\mathrm{max}}_n$ is the aggregated maximum available active power from IBRs according to the MPPT control in the $n$-th area. Constraint \eqref{K^C_lim} ensures adequate deloaded active power and sufficiently large power output during normal operation (without LAAs) to provide desired droop response following the frequency oscillation on both sides triggered by the attack. It is clear that a larger $K^C_{n,n}$ leads to a more stable system but inevitably more operation cost. This trade-off will be balanced in the proposed cyber-resilient economic dispatch model.


\subsection{Recursive Linearization Approach}
\begin{figure}[!t]
    \centering
	\scalebox{0.18}{\includegraphics[]{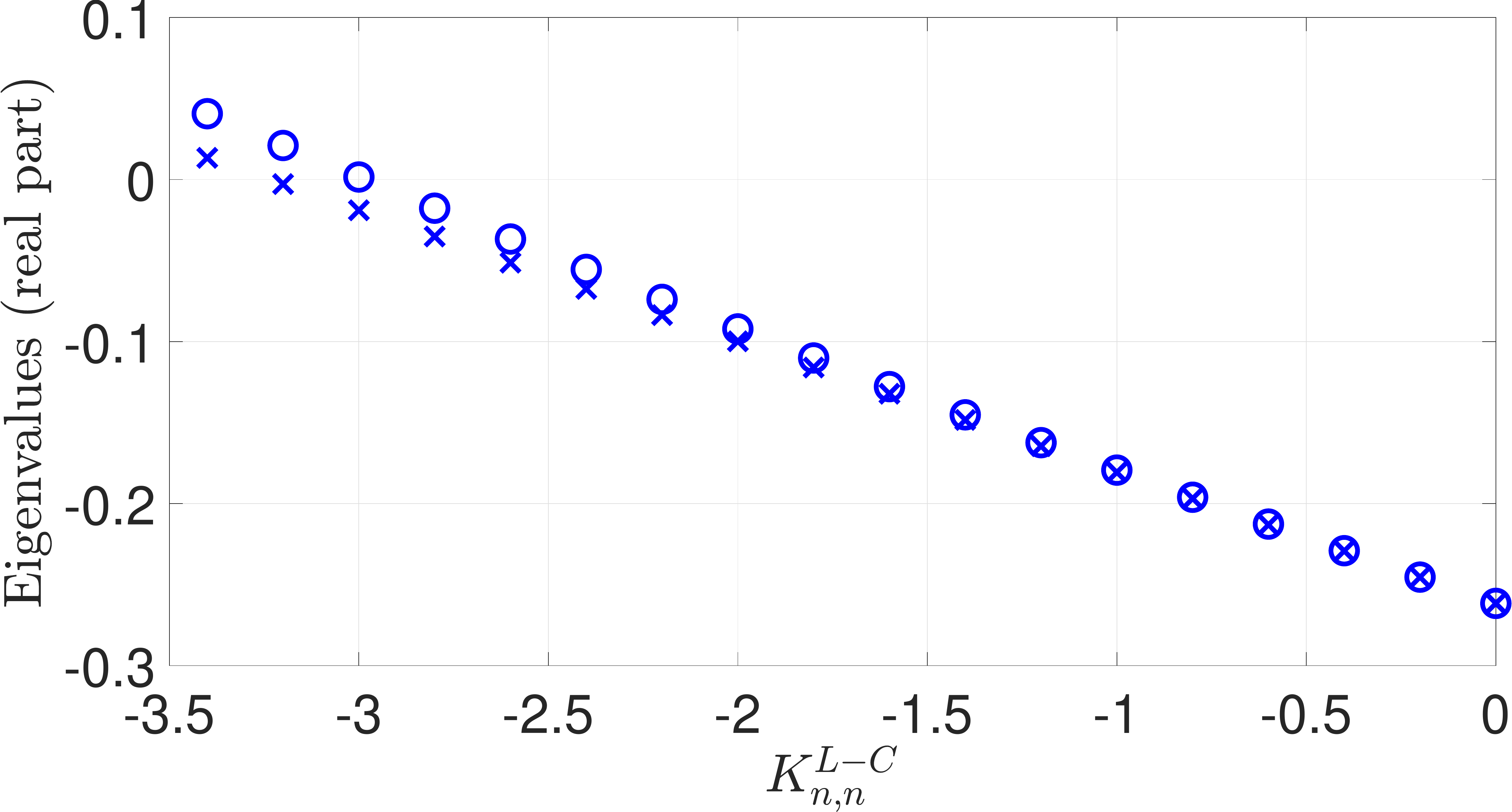}}
    \caption{\label{fig:linear1}Real part of system eigenvalue under single-point LAAs. Circles: true values, Crosses: estimated values.}
\end{figure}

Since the system stability constraint in \eqref{f_sta} is derived based on the sensitivity analysis, it provides a satisfying approximation only if the $K^{L}_{n,n}$ and $K^{C}_{n,n}$ are relatively small. However, this may not always be the cases in reality. To illustrate this issue, first rewritten \eqref{f_sta} giving the relationship in \eqref{sensitivity}:
\begin{align}
\label{f_sta_1}
    \Re{(\lambda_i^0)} + \sum_{n\in\mathcal{N}}  \Re{\left(\frac{\partial\lambda_i^0}{\partial K^L_{n,n}}\right)}(\underbrace{K^L_{n,n} - K^C_{n,n}}_{K^{L-C}_{n,n}})& <0, \nonumber\\
    \forall i &= 1,..., 2N.
\end{align}
An example demonstrating the linearization performance based on \eqref{f_sta_1} is shown in Fig.~\ref{fig:linear1} with only one eigenvalue plotted for simplicity. It is observed that as $K^{L-C}_{n,n}$ decreases from zero, the real part of the eigenvalue increases, which results in an unstable system if it becomes positive. However, the error between the true values and the approximations based on \eqref{f_sta_1} gradually increases in this process due to the larger distance to the linearization point ($K^{L-C}_{n,n}=0$). In addition, since the system stability depends on the sign of the eigenvalues, good approximations should always be ensured for all $K^{L-C}_{n,n}$ that renders eigenvalues with negative and slightly positive real parts, e.g., $[-3.5, 0]$ in this case. Therefore, the approximations in Fig.~\ref{fig:linear1} give unsatisfying results because of the errors around zero. Although this may not occur for certain system parameters, there is no systematic guarantee based on the formulation in \eqref{f_sta_1}.

\begin{algorithm}[!t]
\caption{Recursive Linearization Algorithm}
\label{alg:Linearization}
\begin{algorithmic}[1]
    \State Set $m=0$, $l=0$, $\varepsilon^{m}=0$, $\Delta \phi_{i,n}^{(l)}=0$ and $\mathcal{M}_{i,n}=\{m\}$
    \State Determine the sensitivity range, $\phi_{i,n}=[\underline \phi_{i,n},0]$ 
    \While {$\Delta \phi_{i,n}^{(l)}\ge\underline \phi_{i,n}$}
        \State Obtain the new linearization point, $\Delta \phi_{i,n}^{m}=\Delta \phi_{i,n}^{(l)}$
        \State Calculate the eigenvalue at this point, 
        
        $\lambda_{i,n}^m = \lambda_i\rvert_{K^{L-C}_{n,n}=\Delta \phi_{i,n}^{m}}$
        \State Calculate the sensitivity at this point, $\frac{\partial \lambda_{i,n}^{m}}{\partial K_{n,n}^{L}}$
        \While {$\varepsilon^{m}\le\varepsilon^{\mathrm{lim}}$}
        
            \State $l=l+1$
            \State Move along the sensitivity range,
        
            $\qquad \Delta \phi_{i,n}^{(l)}=\Delta \phi_{i,n}^{(l-1)}-\varepsilon_\phi$
            \State Calculate the true eigenvalue, $\lambda_{i,n}^{(l)}$
            \State Calculate the estimated eigenvalue, $\hat\lambda_{i,n}^{m,(l)}$
            \State Compute the approximation error,
            
            $\qquad\varepsilon^{m}=\left|\Re{(\lambda_{i,n}^{(l)})}-\Re{(\hat\lambda_{i,n}^{m,(l)})}\right|$
        \EndWhile
        \State $\mathcal{M}_{i,n}=\mathcal{M}_{i,n}\cup \{m \}$
        \State $m = m+1$
        \State Reset the approximation error, $\varepsilon^{m}=0$
	\EndWhile
	\State Return $\mathcal{M}_{i,n}$,  $\boldsymbol{\varphi}_{i,n}$ and $\boldsymbol{\partial \varphi}_{i,n}$
\end{algorithmic}
\end{algorithm}
In order to address this problem, a piecewise decision-dependent recursive linearization approach is proposed with the detailed steps shown in Algorithm~\ref{alg:Linearization}. After the initialization, the sensitivity range $\phi_{i,n}$, within which an accurate approximation is desired, is determined first. For instance, $\phi_{i,n}=[-3.5, 0]$ in the case of Fig.~\ref{fig:linear1}. Gradually vary the auxiliary $\Delta \phi_{i,n}^{(l)}$ from $0$ to $\underline \phi_{i,n}$ with a pre-defined fixed step, $\varepsilon_\phi$. At each step, calculate the true eigenvalue, $\lambda_{i,n}^{(l)}$, by setting $K^{L-C}_{n,n}=\Delta \phi_{i,n}^{(l)}$ and the estimated eigenvalue, $\hat\lambda_{i,n}^{m,(l)}$, based on the sensitivity evaluated at the latest linearization point $K^{L-C}_{n,n} = \Delta \phi_{i,n}^{m}$, which are defined as follows:
\begin{subequations}
\begin{align}
    \lambda_{i,n}^{(l)} &= \lambda_i\rvert_{K^{L-C}_{n,n}=\Delta \phi_{i,n}^{(l)}}\\
    \hat\lambda_{i,n}^{m,(l)}& = \lambda_{i,n}^m+\frac{\partial \lambda_{i,n}^{m}}{\partial K_{n,n}^{L}}(\Delta \phi_{i,n}^{(l)}-\Delta \phi_{i,n}^{m}),
\end{align}
\end{subequations}
where $\lambda_{i,n}^m$ is the $i$-th eigenvalue of the system with potential attack or IBR droop control in area $n$, evaluated at the $m$-th linearization point (Step 5), and $\frac{\partial \lambda_{i,n}^{m}}{\partial K_{n,n}^{L}}$ is the associated sensitivity based on \eqref{sensitivity}.

As long as the approximation error is smaller than a pre-specified limit $\varepsilon^{\mathrm{lim}}$, the current linearization point and the associated sensitivity are acceptable. However, once the error becomes larger than the limit, reset the approximation error. Then, choose a new linearization point by setting $\Delta \phi_{i,n}^{m}=\Delta \phi_{i,n}^{(l)}$ and calculate the eigenvalue and the sensitivity at this point. This process is repeated until all the values in the sensitivity range are evaluated (in a discrete manner). To formulate the system stability constraints, all the linearization points and the associated sensitivities are collected:
\begin{subequations}
\begin{align}
    \boldsymbol{\varphi}_{i,n}& = \left\{\left.\phi_{i,n}^m \right| \forall\, m\in \mathcal{M}_{i,n}\right\} \\
    \boldsymbol{\partial \varphi}_{i,n} &= \left\{\left.\frac{\partial \lambda_{i,n}^{m}}{\partial K_{n,n}^{L}}\right|  \forall\, m\in \mathcal{M}_{i,n}\right\}.
\end{align}
\end{subequations}
Note that theoretically this algorithm applies for all $i=1,...,2N$ and $n\in \mathcal{N}$. However, in practice, not all the eigenvalues and their sensitivities with respect to all the areas are needed. Given the system information and the prediction of the potential attacks, it is very likely that the numbers of vulnerable eigenvalues and sensitives are limited. Those eigenvalues that are barely influenced by the attack and the IBR droop control, thus never entering the right-half plane, can be neglected.
\begin{figure}[!t]
    \centering
	\scalebox{0.18}{\includegraphics[]{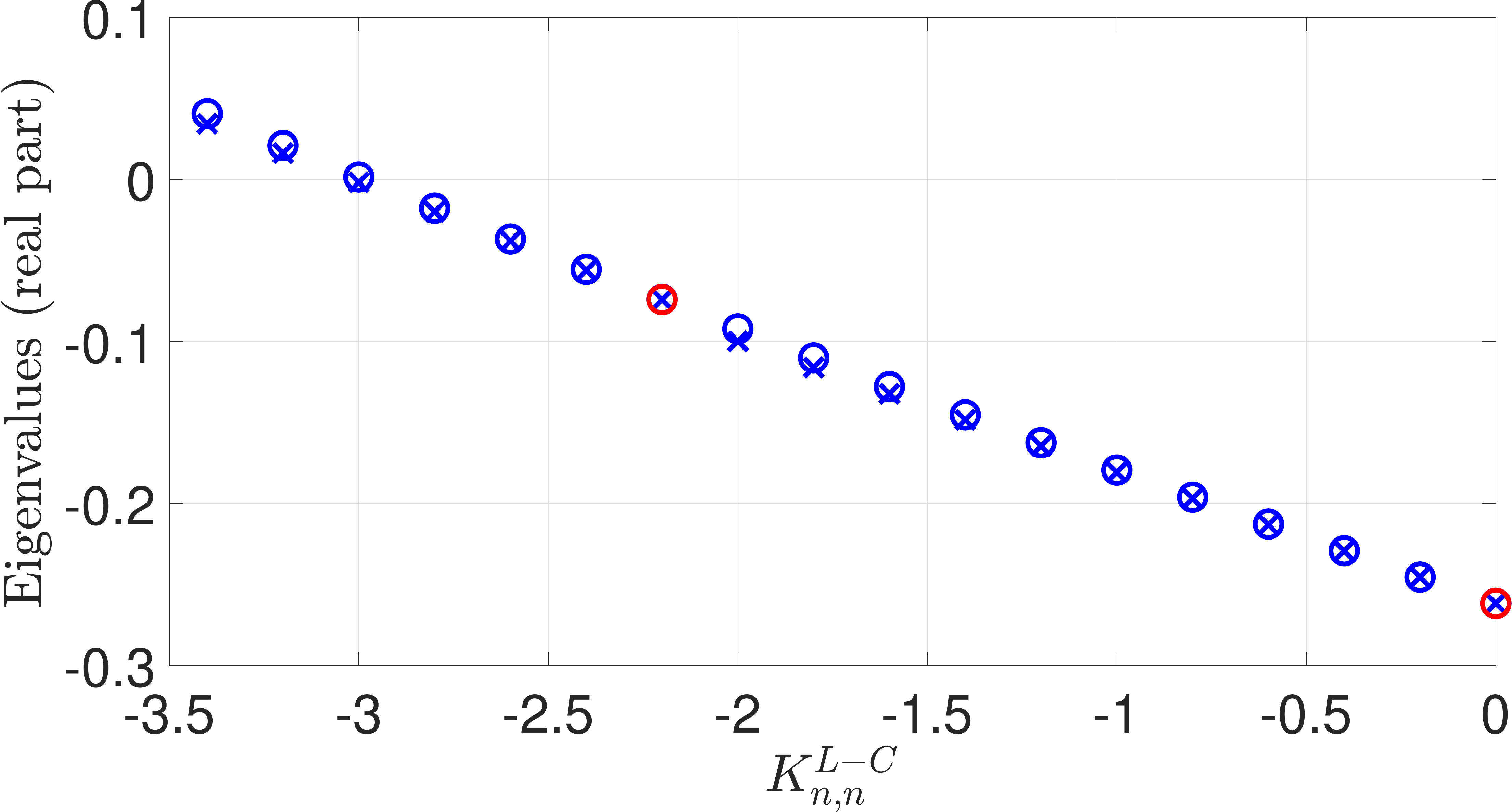}}
    \caption{\label{fig:linear2}Real part of system eigenvalue using proposed algorithm. Circles: true values, Crosses: estimated values.}
\end{figure}

An example of the linearization performance applying Algorithm~\ref{alg:Linearization} is depicted in Fig.~\ref{fig:linear2} where the red circles represent the linearization point, i.e., $\phi_{i,n}^m \in \{0,-2.2\}$ in this case. It is observed that the gradually increasing approximation error due to the deviation from the linearization point is reset at $K_{n,n}^{L-C} = -2.2$, where a new linearization point is defined. Therefore, with the proposed recursive linearization approach, any desired accuracy, specified by $\varepsilon^{\mathrm{lim}}$, can be achieved within the entire sensitivity range.

\subsection{System Stability Constraint Reformulation}
The system stability constraint \eqref{f_sta} which is derived based on sensitivities evaluated at one single linearization point needs to be adapted considering Algorithm~\ref{alg:Linearization}:
\begin{align}
\label{f_sta_2}
    \Re{\left(\lambda_i^0 + \sum_{n\in\mathcal{N}}  \mathcal{F}_n\left(K_{n,n}^{L-C}\right)\right)} <0, \,\forall i = 1,..., 2N,
\end{align}
where $\mathcal{F}_n(\cdot)$ is the eigenvalue variation due to the attack or IBR droop control in the $n$-th area. It is defined as follows:
\begin{align}
\label{Fn_conditional}
    \mathcal{F}_n = 
    \begin{cases}
    {\frac{\partial\lambda_i^0}{\partial K^L_{n,n}}}{K^{L-C}_{n,n}}, & \hspace{-1.5cm} \text{if } K^{L-C}_{n,n}\in (\phi_{i,n}^1, \phi_{i,n}^0]\\
    ...\\
    {\frac{\partial\lambda_i^m}{\partial K^L_{n,n}}}(K^{L-C}_{n,n}-\phi_{i,n}^m)\\
    \quad+ (\lambda_i^m-\lambda_i^0), & \hspace{-1.5cm} \text{if } K^{L-C}_{n,n}\in (\phi_{i,n}^{m+1}, \phi_{i,n}^m] \quad.\\
    ...\\
    {\frac{\partial\lambda_i^{|\mathcal{M}_{i,n}|-1}}{\partial K^L_{n,n}}}(K^{L-C}_{n,n}-  \phi_{i,n}^{|\mathcal{M}_{i,n}|-1})\\
    \quad+ (\lambda_i^{|\mathcal{M}_{i,n}|-1} -\lambda_i^0), & \hspace{-1.5cm} \text{if } K^{L-C}_{n,n}\in [\underline\phi_{i,n}, \phi_{i,n}^{|\mathcal{M}_{i,n}|-1}]
    \end{cases}
\end{align}
Depending on the range of $K^{L-C}_{n,n}$, the sensitivity of the nearest linearization point on its right is used. However, since $K^{L-C}_{n,n}$ is essentially a decision variable, \eqref{f_sta_2} is a decision-dependent constraint, which cannot be incorporated into the optimization problem directly. To address this problem, $|\mathcal{M}_{i,n}|$ binary variables, $\left\{\left.z_{i,n}^{m} \right| m\in \mathcal{M}_{i,n}\right\}$, are introduced for each $i$ and $n$ to indicate to which interval $K_{n,n}^{L-C}$ belongs:
\begin{align}
    \label{z_i,n^m}
    z_{i,n}^{m} & = 
    \begin{cases}
        1 &\text{if}\;\; \phi_{i,n}^{m+1} < K_{n,n}^{L-C} \le  \phi_{i,n}^m \\
        0 & \mathrm{otherwise}.
    \end{cases}
\end{align}
Equation \eqref{z_i,n^m} can be rewrite in the following form by defining ancillary binary variables $z_{i,n}^{m1},\,z_{i,n}^{m2},\,\forall m \in \mathcal{M}_{i,n}$:
\begin{subequations}
\label{z_n12}
    \begin{align}
    z_{i,n}^{m1} & = 
    \begin{cases}
        1 &\text{if}\;\;\phi_{i,n}^{m+1} < K_{n,n}^{L-C} \\
        0 & \mathrm{otherwise}
    \end{cases}\\
    z_{i,n}^{m2} & = 
    \begin{cases}
        1 &\text{if}\;\;K_{n,n}^{L-C} \le \phi_{i,n}^m \\
        0 & \mathrm{otherwise}
    \end{cases}\\
    z_{i,n}^{m} & = z_{i,n}^{m1}  + z_{i,n}^{m2} -1.
    \end{align}
\end{subequations}
As a result, the conditional constraints \eqref{z_n12} can be transformed into linear constraints, $\forall n\in \mathcal{N}$:
\begin{subequations}
    \label{z_n12_linear}
    \begin{align}
    & 0 \le \phi_{i,n}^{m+1} - K_{n,n}^{L-C} + \mathbf{M} z_{i,n}^{m1} < \mathbf{M}
    \\
    & 0 < K_{n,n}^{L-C} - \phi_{i,n}^{m} +\mathbf{M} z_{i,n}^{m2} \le \mathbf{M}
    \\
    & z_{i,n}^{m} = z_{i,n}^{m1}  + z_{i,n}^{m2} -1 \\
    & z_{i,n}^{m},\, z_{i,n}^{m1},\, z_{i,n}^{m2}\,\in \{0,1\}
    \end{align}
\end{subequations}
where $\mathbf{M}>0$ is a sufficiently large constant. With the binary variables $z_{i,n}^{m}$, the conditional constraints \eqref{Fn_conditional} can be converted to:
\begin{equation}
\label{Fn}
    \mathbf{F}_n = \sum_{m\in \mathcal{M}_{i,n}} \mathbf{F}_n^m z_{i,n}^{m},
\end{equation}
where
\begin{equation}
\label{F_n^m}
    \mathbf{F}_n^m =  {\frac{\partial\lambda_i^m}{\partial K^L_{n,n}}}(K^{L-C}_{n,n}-\phi_{i,n}^m)
    + (\lambda_i^m-\lambda_i^0), \,\,\forall m \in \mathcal{M}_{i,n}.
\end{equation}
The only nonlinearity in \eqref{Fn}, \eqref{F_n^m} is the term $K^{L-C}_{n,n}z_{i,n}^{m}$, which is the product of a continuous and a binary variable. This can be easily linearized through the standard Big-M method, which is not covered here. Finally, the system stability constraint under LAAs and the IBR droop control can be obtained by combining \eqref{f_sta_2} with $\mathcal{F}_n$ being replaced by $\mathbf{F}_n$ together with \eqref{z_n12_linear}, \eqref{Fn} and \eqref{F_n^m}:
\begin{align}
\label{f_sta_3}
    \Re\bigg(\lambda_i^0 & + \sum_{n} \bigg ( \sum_{m} \Big(\frac{\partial\lambda_i^m}{\partial K^L_{n,n}}(K^{L-C}_{n,n}-\phi_{i,n}^m) \nonumber\\
    &+(\lambda_i^m-\lambda_i^0)\Big) z_{i,n}^{m}\bigg)\bigg)  <0, \,\forall i = 1,..., 2N.
\end{align}


\section{Incorporating Uncertainty of Attack  Detection Into CRED And Overall System Operation} \label{sec:4}
{\color{black} The framework presented thus far assumes perfect knowledge of the attack parameters $K^L$. In real-world operation, attack mitigation operation such as CRED follows the attack detection/attack parameter estimation phase. Existing works \cite{AminiIdentification2019, lakshminarayana2021datadriven} adopt a data-driven approach for this purpose using frequency/phase angle data monitored by phasor measurement units. However, due to the stochastic nature of the detection algorithms (e.g., machine learning based detectors proposed in \cite{lakshminarayana2021datadriven}), and factors such as sensor measurement noise, the predicted attack parameters incur  estimation  errors. In this section, we present a framework to incorporate such errors into the CRED algorithm and provide details on how  CRED can be incorporated into the overall system operation.}





\subsection{Distributionally Robust Chance Constraint}


In order to account for the uncertainty associated with the estimation of LAAs parameters, the system stability constraints are further reformulated into distributionally robust form. Assume that the first and second moment of LAAs gains are known whereas the detailed distribution is unknown. {\color{black} For instance, the aforementioned first and second moments can be computed numerically by analyzing data traces obtained from the attack detection/parameter estimation process (more details are provided in the Section~\ref{sec:5}).}

Therefore, the random variable $K^L_{n,n},\, \forall n\in \mathcal{N}$ can be modeled by the following ambiguity set:
\begin{align}
\label{ambiguity_K^L}
    \mathcal{P} =\Big\{\mathbf{D} \in \Phi(K^L_{n,n}):\; & \mathrm{E}^\mathbf{D}(K^L_{n,n}) = \bar K^{L}_{n,n},\nonumber \\
    &\mathrm{Var}^\mathbf{D}(K^L_{n,n}) =\sigma_n^2 \Big\},
\end{align}
where $\Phi(K^L_{n,n})$ denotes the set of all probability distributions on $K^L_{n,n}$; $\bar K^{L}_{n,n}$ and $\sigma_n^2$ denote the mean and variation of $K^{L}_{n,n}$. In addition, assume that the attacks in different areas are independent with each other, i.e., 
\begin{equation}
    \mathrm{cov}\left(K^{L}_{n,n}, K^{L}_{m,m}\right)=0,\,\;\forall n,m\in \mathcal{N}, n\neq m.
\end{equation}
By modifying \eqref{f_sta_3}, the distributionally robust chance constraint of system stability, $\forall i = 1,..., 2N$, can be thus written as:
\begin{align}
\label{f_sta_DR}
    \inf_{\mathbf{D}\in\mathcal{P}} \mathrm{Pr} \Bigg\{ \Re\bigg(\lambda_i^0 & + \sum_{n} \bigg ( \sum_{m} \Big(\frac{\partial\lambda_i^m}{\partial K^L_{n,n}}(K^{L}_{n,n}-K^{C}_{n,n}-\phi_{i,n}^m) \nonumber\\
    &+(\lambda_i^m-\lambda_i^0)\Big) z_{i,n}^{m}\bigg)\bigg)  <0\Bigg\} \ge \eta,
\end{align}
where $\eta\in (0,1)$ is the confidence level specified by system operators. According to \textit{Theory 3.1} in \cite{calafiore_ghaoui_2022}, the above distributionally robust chance constraint is equivalent to:
\begin{subequations}
\label{stability_ctr}
\begin{align}
    \Re\bigg(\lambda_i^0 & + \sum_{n} \bigg ( \sum_{m} \Big(\frac{\partial\lambda_i^m}{\partial K^L_{n,n}}(\bar K^{L}_{n,n} + k_\eta \sigma_n \nonumber\\
    & -K^{C}_{n,n}-\phi_{i,n}^m) +(\lambda_i^m-\lambda_i^0)\Big) z_{i,n}^{m}\bigg)\bigg)  <0 \\
    & \hspace{1.96cm} k_\eta = \sqrt{\frac{\eta}{1-\eta}},
\end{align}
\end{subequations}
which ensures system stability given certain confidence level.
\subsection{Overall System Operation Framework} \label{sec:4.2}
\begin{figure}[!t]
    \centering
	\scalebox{0.75}{\includegraphics[]{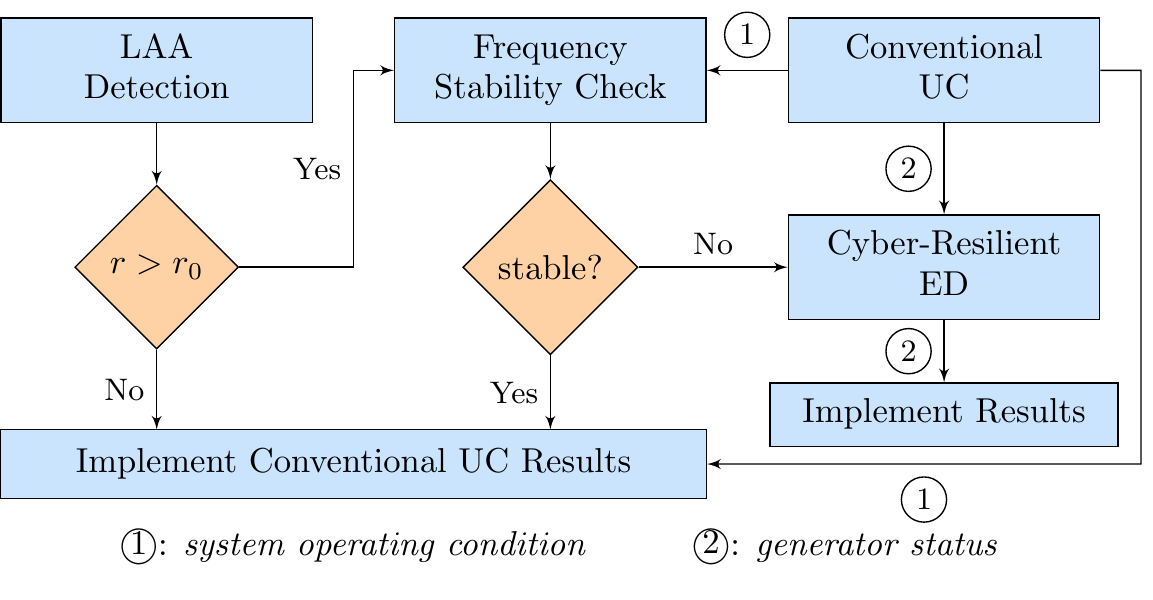}}
    \caption{\label{fig:Framework}General framework of the proposed model in system operation.}
\end{figure}
The overall framework of the proposed CRED is integrated in the system scheduling model as illustrated in Fig.~\ref{fig:Framework}. Since the LAAs are low frequency high impact events, it is economically inefficient to run the CRED algorithm during normal system operation. Instead, at this stage, conventional Unit Commitment (UC) with N-1 frequency security constraints (maximum RoCoF, frequency nadir and steady-state frequency) as proposed in \cite{9066910}, is conducted and the LAA detection is constantly performed. The objective of the conventional UC problem is to minimize the expected cost over all nodes in the given scenario tree:
\begin{equation}
    \label{eq:SUC}
    \min \sum_{n\in \mathcal{N}} \pi (n) \left( \sum_{g\in \mathcal{G}}  C_g(n) + \Delta t(n) c^s P^s(n) \right)
\end{equation}
where $\pi(n)$ is the probability of scenario $n\in \mathcal{N}$ and $C_g(n)$ is the operation cost of unit $g\in \mathcal{G}$ in scenario n including startup, no-load and marginal cost; $\Delta t(n)c^sP^s(n)$ represents the cost of the load shedding in scenario n with the three terms being the time step of scenario n, load shedding cost and shed load. The objective function \eqref{eq:SUC} is subjected to a number of constraints. Due to the space limitation, all the conventional UC constraints such as those related to thermal generation units, power balance and frequency security are not listed in the paper. The readers can refer \cite{7115982,9066910} for more details regarding the constraints and scenario tree building. Note that at this stage, the system stability constraints are not included in the scheduling model.

However, if a potential LAA is detected, indicated by a pre-defined criterion, $r>r_0$, (e.g., using existing approaches \cite{lakshminarayana2021datadriven}), then the system resilience needs to be guaranteed under this circumstance. The operating conditions obtained from conventional UC are checked first, against system stability given the estimation of LAA. Since the system parameters, operating conditions and the attack information are known, this process can be conducted fast and easily by assessing the eigenvalues of the system \eqref{stability_ctr}, thus not being covered here. A stable system renders no further action and the results from conventional UC are implemented, whereas an unstable system requires CRED algorithm to be implemented in order to ensure stable system operation (until the attack can be potentially isolated) and avoid potential blackouts. 

The CRED takes the input of generator status from conventional UC and outputs the resilient system operating conditions. The optimization model of the CRED remains the same as the conventional UC, except that the generator states are fixed according to the conventional UC and the constraint \eqref{stability_ctr} is added to guarantee the system stability under the detected LAAs. Constraint \eqref{stability_ctr} requires IBR droop control and deloading of RES, which influences generation cost and the optimization objective as defined in \eqref{eq:SUC}. Larger droop gains lead to a more stable system with higher operational cost. This trade-off can be balanced within the proposed CRED framework, ensuring the system stability under LAAs in a most cost-effective manner. Note that if the actual attack parameters (such as attack controller gain $K^L$) cannot be identified accurately, then the system operator can always use the worst-case attack parameters in the implementation of CRED framework, {by replacing \eqref{stability_ctr} with \eqref{f_sta_3}}. Additionally, during the CRED process, we assume that the loss of generation and the LAAs would not occur simultaneously.

\section{Case Studies}\label{sec:5}
In this section, the effectiveness of the proposed CRED algorithm and its impact on system operation are assessed through simulations in the modified IEEE Reliability Test System (72 buses, 33 machines). The system is divided into three areas and the parameters are available in \cite{7339813,780914}. In addition, wind generation is added in Area 2 to realize different IBR penetrations and it is assumed that the attack only occurs in Area 2 since an attack in other areas has little impact on system operation as demonstrated in Section~\ref{sec:5.2}. Other system parameters are set as follows: load demand $P^D\in [4.79, 8.55]\,\mathrm{GW}$, load damping $D = 0.5\% P^D / 1\,\mathrm{Hz}$, base power $P_B = 1\mathrm{GW}$, confidence level $\eta = 95\%$. The frequency limits of nadir, steady-state value and RoCoF are set as: $0.8\,\mathrm{Hz}$, $0.5\,\mathrm{Hz}$ and $0.5\,\mathrm{Hz/s}$ respectively. The weather conditions are obtained from online numerical weather prediction \cite{weather}. We consider an optimization problem with a time horizon of 24 hours and time step of 1 hour, which is solved by Gurobi (8.1.0) on a PC with Intel(R) Core(TM) i7-7820X CPU @ 3.60GHz and RAM of 64 GB. \textcolor{black}{Note that the SI from IBRs are not considered, i.e., $M_s=0$ in \eqref{sys_dyn0}, for all the case studies except in Section~\ref{sec:5.3} where its impact is explicitly investigated.}

\begin{figure}[!t]
  \centering
    \begin{subfigure}{0.485\textwidth}
        \centering
        \scalebox{1.2}{\includegraphics[]{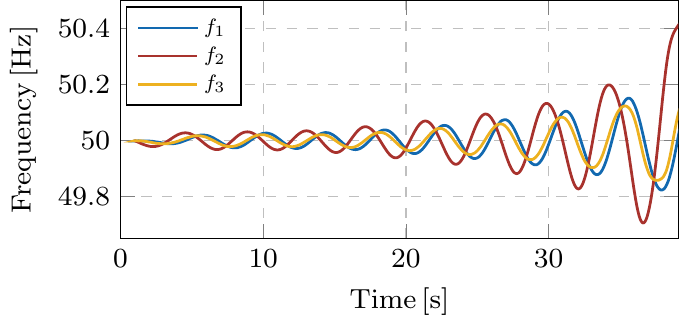}} 
        \vspace{-0.35em}  
        \caption{}
        \label{fig:timeSim_A}       
    \end{subfigure} 
    \begin{subfigure}{0.485\textwidth}
        \centering
        \scalebox{1.2}{\includegraphics[]{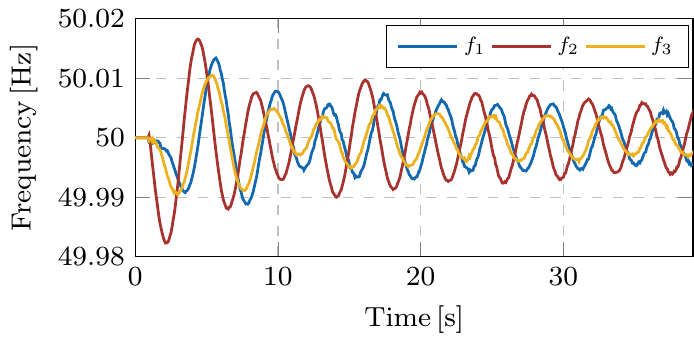}} 
        \vspace{-0.35em}  
        \caption{}
        \label{fig:timeSim_B}       
    \end{subfigure} 
  \caption{\label{fig:plots}System frequency evolution after a small step disturbance at $t=1\,\mathrm{s}$: (a) without CRED; (b) with CRED.}
\end{figure}

\subsection{Time domain simulation} \label{sec:5.0}
The effectiveness of the proposed CRED in terms of maintaining system stability is validated through time domain simulation. Fig.~\ref{fig:plots} depicts the frequency dynamics of the studied system under LAA at Area 2 with and without the CRED, where $f_{i\in\{1,2,3\}}$ represents the averaged frequency of the $i$-th area. The same amount of accessible vulnerable load (30\%) to the attacker is assumed for both situations. In order to assess the system stability, the frequency variations are triggered by applying a small step disturbance at $t=1\,\mathrm{s}$. In Fig.~\ref{fig:timeSim_A}, the system becomes unstable under LAAs without CRED, where the frequencies in all the areas oscillate with increasing magnitude. $f_2$ is spotted with the largest oscillation magnitude as the LAA is applied in this area. This growing frequency deviation would eventually lead to generator trips and even cascade failures, endangering the system operation. On the contrary, with the proposed CRED (Fig.~\ref{fig:timeSim_B}), the LAA with the same magnitude cannot destabilize the system, since there is enough damping in the system and the oscillations due to the LAA gradually attenuate. Note the difference in the frequency range of the two sub-figures.

\begin{table}[!b]
\renewcommand{\arraystretch}{1.2}
\caption{System Parameters w/o and with CRED}
\label{tab:sys_para}
\noindent
\centering
    \begin{minipage}{\linewidth} 
    \renewcommand\footnoterule{\vspace*{-5pt}} 
    \begin{center}
        \begin{tabular}{ c | c | c | c | c | c | c}
            \toprule
            \multirow{2}{3em}{\textbf{CRED}} &$\boldsymbol{K^L}$ &$\boldsymbol{K^C}$ &$\boldsymbol{P_{\mathrm{res}}}$  & $\boldsymbol{P^W}$ & $\boldsymbol{P^G}$ & $\boldsymbol{\mathrm{Cost}}$\\ 
            &$[\mathrm{p.u.}]$ &$[\mathrm{p.u.}]$ &$[\mathrm{GW}]$  & $[\mathrm{GW}]$ & $[\mathrm{GW}]$ & $[\mathrm{k\pounds}]$\\ 
            \cline{1-7}
            $\mathrm{w/o}$ & $26.70$  & $0$  & $0$ & $3.82$ & $4.74$ & $68.73$ \\
            \cline{1-7}
            $\mathrm{with}$ & $26.70$  & $18.85$  & $0.30$ & $3.51$ & $5.04$ & $80.21$ \\
           \bottomrule
        \end{tabular}
        \end{center}
    \end{minipage}
\end{table} 

\textcolor{black}{Other parameters related to the two systems in Fig.~\ref{fig:plots} are listed in Table~\ref{tab:sys_para}, where $K^L$ and $K^C$ are the DLAA and IBR droop gains; $P_{\mathrm{res}}$, $P^W$, $P^G$ and $\mathrm{Cost}$ represent the reserved wind power for IBR droop response, total power from wind and synchronous generation and system operation cost respectively. It is observed that to maintain system stability $0.6\,\mathrm{GW}$ of wind power is deloaded, hence more power from SGs and larger operation cost.}


\subsection{Assessment of CRED with various knowledge of attack parameters} \label{sec:5.1}
\begin{figure}[!t] 
	\centering
	\vspace{-0.4cm}
	\scalebox{1.2}{\includegraphics[]{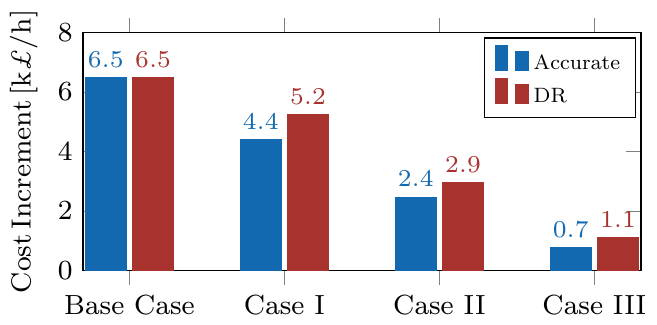}}
	\caption{Cost increment of CRED under different attack knowledge.}
	\label{fig:Estimation}
\end{figure}
The additional system operation cost to ensure the cyber-resilience is shown in Fig.~\ref{fig:Estimation}. Three different situations are considered: (1) worst case assumption of the attack parameter where the system operator has no estimation of the actual attacks and {assumes the attacker has access to all the vulnerable load in the system} (Base Case); (2) accurate estimation of the actual attacks where the real attack gains can be predicted perfectly (blue bars in Case I-III); (3) Distributional Robust (DR) case where only the first and second moments of the attack parameters are assumed to be known based on some attack estimation methods (red bars in Case I-III).  In particular, for (3), we implement the sparse identification of non-linear dynamics (SINDy) algorithm proposed in \cite{lakshminarayana2021datadriven} for LAAs attack parameter estimation (we choose the SINDy algorithm as the algorithm is robust and performs well over a wide range of system parameters) over $1000$ runs, and compute the required first and second moments numerically as shown in Table~\ref{tab:moments}.
Note that the installed wind capacity in Area 2 for all the cases is $5\,\mathrm{GW}$ and the total vulnerable load is $30\%$ of the load in this area. Since attacks with the same magnitude at different hours of a day may have different effect on system operation, the cost increment in this work is the averaged value among the cost increment due to the attack at each hour of the day.

\begin{table}[!b]
\renewcommand{\arraystretch}{1.2}
\caption{Attack Parameters Associated with Fig.~\ref{fig:Estimation}}
\label{tab:moments}
\noindent
\centering
    \begin{minipage}{\linewidth} 
    \renewcommand\footnoterule{\vspace*{-5pt}} 
    \begin{center}
        \begin{tabular}{ c | c | c | c | c }
            \toprule
             $\,$ &$\boldsymbol{\mathrm{Base\, Case}}$ &$\boldsymbol{\mathrm{Case\, I}}$ &$\boldsymbol{\mathrm{Case\, II}}$ & $\boldsymbol{\mathrm{Case\, III}}$ \\ 
            \cline{1-5}
            $\boldsymbol{K^L\,\mathrm{[p.u.]}}$ & $22.31$  & $16.96$  & $13.50$ & $9.11$ \\ 
            \cline{1-5}
            $\boldsymbol{\bar K^L\,\mathrm{[p.u.]}}$ & $22.31$  & $17.59$  & $13.19$ & $8.80$ \\ 
            \cline{1-5}
            $\boldsymbol{\sigma\,\mathrm{[p.u.]}}$ & $0$  & $0.48$  & $0.36$ & $0.24$\\ 
           \bottomrule
        \end{tabular}
        \end{center}
    \end{minipage}
\end{table} 

It can be observed from the figure that Base Case leads to largest cost increment since it assumes all the vulnerable load can be accessed by the attacker to generate the attack, thus being most conservative. In this case, the accurate and DR approaches make no difference. Furthermore, Case I to Case III represent the circumstances where the attacker has $80\%$, $60\%$ and $40\%$ accessibility to the entire vulnerable load respectively. As expected, when the attacker has lower accessibility to the vulnerable load, less additional cost is generated for both accurate estimation and DR approach, to maintain the system stability. In all these cases, the DR approach causes slightly larger cost increment compared with accurate estimation due to the uncertainty management associated with the attack parameter estimation. However, it is also clear that the proposed DR approach can reduce the attack-induced operation cost giving actual estimation of the attack parameters compared with the worst case assumption. 

{Note that the dynamic simulation shown in Fig.~\ref{fig:timeSim_B} corresponds to the accurate attack estimation case. Since the DR case presents a very similar trend except a slightly faster attenuation rate due to the larger damping, it is not included.}

\subsection{Impact of wind penetration on LAAs and CRED} \label{sec:5.2}
\begin{figure}[!t]
	\centering
	\vspace{-0.4cm}
	\scalebox{1.2}{\includegraphics[]{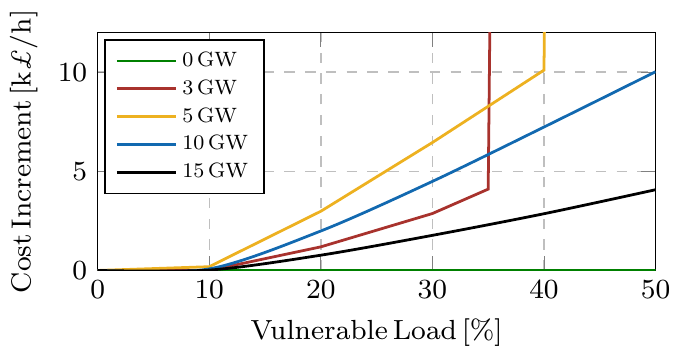}}
	\caption{Impact of wind penetration and vulnerable load on system operation cost considering potential LAAs.}
	\label{fig:Cost_Wind_Attack}
\end{figure}
In conventional power system where the main power sources are SGs, the LAAs may not be able to destabilize the system as there is sufficient damping in the system. However, the consequence of LAAs could become more obvious with higher wind penetration where less damping from SGs is available. To demonstrate this effect, Fig.~\ref{fig:Cost_Wind_Attack} depicts system cost increment due to the attack with increasing percentage of vulnerable load, where each line represents a different installed wind capacity. 

If no wind generation is installed in the system (green line), the LAAs do not influence the system operation cost because the SGs are capable of maintaining the system stability even if $50\%$ of the load in Area 2 is vulnerable. Furthermore, as the installed wind capacity increases to $3\,\mathrm{GW}$ and $5\,\mathrm{GW}$, less SGs are dispatched in the system, making the system less stable. Therefore, additional damping ($K^C$) is required from IBRs to maintain system stability under the LAAs, which results in the deloading of wind power and more dispatched power from SGs, hence more operation cost. A larger wind penetration and higher vulnerable load make the above effect more evident, corresponding to higher cost increment. Notably, after the vulnerable load increases to about $35\%$ and $40\%$ in the cases of $3\,\mathrm{GW}$ and $5\,\mathrm{GW}$ respectively, the system operation cost grows dramatically. This is because there is not enough wind power to be deloaded for more damping provision and some of the load has to be curtailed to maintain system stability, thus generating significant load shedding cost. 

However, as the wind penetration continues increasing (blue and black curves), we notice that the cost increment due to the attack starts to decrease, since to ensure the N-1 frequency security constraints, certain amount of SGs has to remain connected in the system and more wind power is available to supply additional damping, making it easier to maintain system stability. Moreover, the substantial cost increase induced by the load shedding disappears with the sufficient wind power.

\begin{figure}[!t] 
	\centering
	\vspace{-0.4cm}
	\scalebox{1.1}{\includegraphics[]{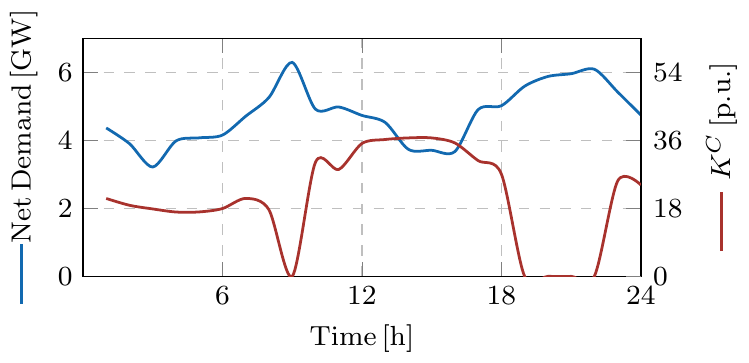}}
	\caption{Relationship between net demand and $K^C$.}
	\label{fig:Net_D}
\end{figure}

An example of the additional damping from IBRs ($K^C$) under the attack at different hours in the day, and its relationship with the net demand \textcolor{black}{($P^D-P^W$)} in the system are depicted in Fig.~\ref{fig:Net_D}, {where the vulnerable load is $30\%$ of the total load and the wind capacity is $5\,\mathrm{GW}$}. A clear negative correlated relationship can be spotted, i.e., higher net demand leads to less $K^C$ from wind generation. This is due to the reason that more SGs are dispatched online to supply the higher net demand, hence less damping from IBRs are needed to maintain system stability. {Note that the equivalent damping from SGs is around $20\,\mathrm{p.u.}$ based on their own capacities \cite{7339813}.} 

\subsection{Impact of synthetic inertia provision on CRED} \label{sec:5.3}
\begin{figure}[!b] 
	\centering
	\vspace{-0.4cm}
	\scalebox{1.2}{\includegraphics[]{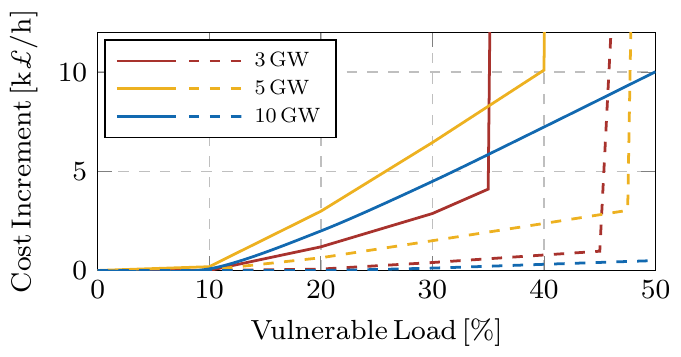}}
	\caption{Impact of synthetic inertia provision on CRED. Solid lines: without SI, Dashed lines: with SI.}
	\label{fig:Cost_SI_Attack}
\end{figure}
During the conventional UC process in Fig.~\ref{fig:Framework}, the N-1 frequency security constraints are considered such that the constraints of frequency metrics will not be violated. The previous section (Sec.~\ref{sec:5.2}) only utilizes the inertia and primary frequency responses from SGs to ensure the frequency security. Alternatively, fast frequency responses from IBRs can also help to maintain the frequency constraints. Specifically, we consider the Synthetic Inertia (SI) provision from wind turbines \cite{9066910} and its influences on the LAAs and the CRED are analyzed here. {Notably, since SI is provided using the stored kinetic energy of wind turbines, there is no additional cost due to SI provision.}

The results are illustrated in Fig.~\ref{fig:Cost_SI_Attack}, where cost increments with various wind penetration are considered. The solid and dashed lines represent the situation without and with SI provision respectively. Compared with the previous situation (without SI), once SI is incorporated into the convention UC process, the additional operation cost to ensure the system resilience under LAAs reduces significantly at different wind penetrations. This is due to the fact that the synthetic inertia from the wind turbines in Area 2 decreases the eigenvalue sensitivities with respect to the attacks, making the system more stable. Therefore, attacks with the same magnitude as in the no SI cases have less effects on destabilizing the system, thus less cost increment. For the same reason, the percentage of vulnerable load after which load shedding is needed to ensure system stability in the cases with $3\,\mathrm{GW}$ and $5\,\mathrm{GW}$ wind power also increases by about $10\%$ and $8\%$ respectively.

\subsection{CRED under different battery energy storage conditions} \label{sec:5.4}
\begin{figure}[!t] 
	\centering
	\vspace{-0.4cm}
	\scalebox{1.2}{\includegraphics[]{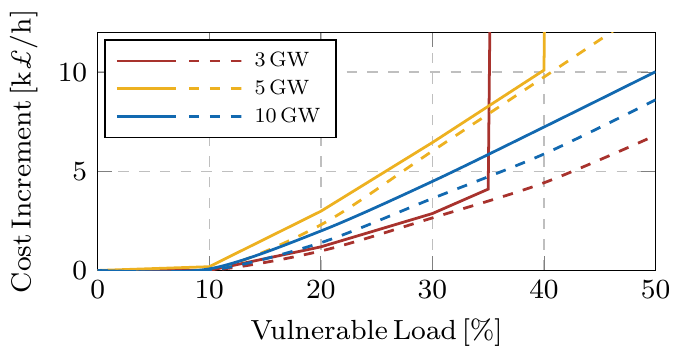}}
	\caption{Impact of energy storage on CRED. Solid lines: without storage, Dashed lines: with storage.}
	\label{fig:Cost_Storage_Attack}
\end{figure}

\begin{table}[!b]
\renewcommand{\arraystretch}{1.2}
\caption{Parameters of Battery Storage Devices}
\label{tab:battery_attack}
\noindent
\centering
    \begin{minipage}{\linewidth} 
    \renewcommand\footnoterule{\vspace*{-5pt}} 
    \begin{center}
        \begin{tabular}{ c | c | c | c | c }
            \toprule
             $\boldsymbol{\mathrm{SoC}_{\mathrm{min}}}$ &$\boldsymbol{\mathrm{SoC}_{\mathrm{max}}}$ &$\boldsymbol{\eta_b}$ &$\boldsymbol{|\bar P_b|\,\mathrm{[GW]}}$  & $\boldsymbol{E_b\,\mathrm{[GWh]}}$ \\ 
            \cline{1-5}
            $20\%$ & $80\%$  & $0.9$  & $5$ & $15$ \\ 
           \bottomrule
        \end{tabular}
        \end{center}
    \end{minipage}
\end{table} 

In power systems with high wind penetration, it is often the situation that certain amount of battery energy storage is preferred to better utilize the renewable energy and mitigate the uncertainty in the system. This subsection investigates how the battery energy storage would influence the resilience of the system under LAAs. Fig.~\ref{fig:Cost_Storage_Attack} presents the system operation cost increment to maintain system stability under LAAs without (solid curves) and with (dashed curves) battery energy storage. The parameters of battery energy storage are shown in Table~\ref{tab:battery_attack}, with $\mathrm{SoC}_{\mathrm{min}}$, $\mathrm{SoC}_{\mathrm{max}}$, $\eta_b$, $|\bar P_b|$ and $E_b$ denote the minimum, maximum state of charge, efficiency, maximum (dis)charging power and energy capacity. It is observed that the overall impact of battery energy storage on the system operation against the LAAs is not very remarkable for most of the cases in Fig.~\ref{fig:Cost_Storage_Attack}, where there is no load shedding. This is because the energy usages with battery storage in the system has already been optimized in the conventional UC problem and once the attack is detected, the additional cost due to the power re-dispatch would not change considerably compared with the case where no storage is installed in the system. Hence, the dashed lines are slightly below the solid lines with the same color. However, the load shedding in the cases of $3\,\mathrm{GW}$ and $5\,\mathrm{GW}$ wind capacity disappears after including battery energy storage since more damping can be supplied by the storage system when the wind power is insufficient, therefore no need of load shedding. 

It should be noted that although battery energy storage system could reduce the total operation cost for balancing generation and demand, it is not reflected in this figure since the cost increment is always obtained by comparing with the case where no attack would occur.

\section{Conclusion and Future Work} \label{sec:6}
A cyber-resilient economic dispatch model is proposed in this paper. 
The effectiveness of the proposed CRED is demonstrated through dynamic simulations. Case studies also illustrate the benefit and essential to consider the LAA parameter estimation. The LAAs have larger impact on system stability in systems with high wind penetration. The operation cost increment to maintain system stability is also influenced by the SI provision from wind turbines and the energy storage in the system. As a follow-up of this work we are looking at 1) the impact of time delay in attack detection and using more advanced and flexible defense strategies and 2) the influence of varying the droop gain of IBRs on the converter-driven instability in weak grids.


\bibliographystyle{IEEEtran}
\bibliography{bibliography}
\end{document}


\begin{varwidth}{\linewidth}

\begin{tikzpicture}
\begin{axis}[
    scaled ticks=false,
    tick label style={/pgf/number format/fixed},
    colormap name=viridis,
    width=7.25cm,
    height=4cm,
    xlabel={$\mathrm{Vulnerable\,Load\,[\%]}$},
    ylabel={$\mathrm{Cost\, Increment\,[k\pounds/h]}$},
    xmin=0, xmax=50,
    ymin=0, ymax=12,
    xmajorgrids=true,
    ymajorgrids=true,
    legend style={at={(axis cs:0.6, 11.65)},anchor=north west,nodes={scale=0.75, transform shape}, legend columns=2},
    legend cell align={left},
    grid style=dashed,
]
\footnotesize
\addplot[
    thick,
    color=pRed,
    ]
    table {data/Cost_Wind/data2.txt};        
    \addlegendentry{\footnotesize }
\addplot[
    thick,
    color=pRed,
    dashed,
    ]
    table {data/SI/data0.txt};        
    \addlegendentry{\footnotesize  $3\,\mathrm{GW}$}
\addplot[
    thick,
    color=pYellow,
    ]
    table {data/Cost_Wind/data3.txt};        
    \addlegendentry{\footnotesize }
\addplot[
    thick,
    dashed,
    color=pYellow,
    ]
    table {data/SI/data1.txt};     
    \addlegendentry{\footnotesize  $5\,\mathrm{GW}$}
\addplot[
    smooth,
    thick,
    color=pBlue,
    ]
    table {data/Cost_Wind/data4.txt};        
    \addlegendentry{\footnotesize  }
\addplot[
    smooth,
    thick,
    dashed,
    color=pBlue,
    ]
    table {data//SI/data2.txt};        
    \addlegendentry{\footnotesize  $10\,\mathrm{GW}$}
\end{axis}

\end{tikzpicture} 

\end{varwidth}